\documentclass[conference]{IEEEtran}

\usepackage{style}
\IEEEoverridecommandlockouts

\begin{document}

\title{SAFE ma-QAOA: Surrogate-Assisted and Fine-Tuning Enhanced Multi-Angle QAOA with Parameter Distillation}

\author{%
\IEEEauthorblockN{Hyunwoo Kim\textsuperscript{*}}
\IEEEauthorblockA{\textit{NORMA, Inc.} \\
Seoul, Republic of Korea \\
hw\_kim@norma.co.kr}
\and
\IEEEauthorblockN{Youngseok Lee\textsuperscript{*}}
\IEEEauthorblockA{\textit{NORMA, Inc.} \\
Seoul, Republic of Korea \\
ys\_lee@norma.co.kr}
\thanks{\textsuperscript{*}These authors contributed equally to this work.}
}

\maketitle

\begin{abstract}
The multi-angle Quantum Approximate Optimization Algorithm (ma-QAOA) extends the Quantum Approximate Optimization Algorithm (QAOA) by assigning a larger number of independent variational parameters,
thereby increasing expressivity and improving performance at low circuit depths.
However, this larger parameterization makes training more difficult and requires repeated circuit evaluations for gradient-based optimization.
In this work, we propose the Surrogate-Assisted and Fine-tuning Enhanced (SAFE) framework. SAFE first uses Low-Weight Pauli Propagation (LWPP) as a classical surrogate for pre-training ma-QAOA parameters before exact optimization.
SAFE then applies parameter distillation, which removes angles that remain near zero after surrogate pre-training.
Finally, SAFE performs exact fine-tuning by optimizing the remaining active parameters using the exact energy objective.
We evaluate SAFE on instances of the Sherrington-Kirkpatrick model, two-dimensional square-lattice spin glass, and Max-Cut.
SAFE with distillation provides the strongest overall results relative to exact-only: (i) a 64.3 percent reduction in active parameter count
and (ii) a 94.5 percent reduction in estimated QPU workload.
Within the SAFE workflow, adding distillation further reduces the number of optimizer steps required to reach the near-optimal regime by 44.4 percent relative to SAFE without distillation.
These results provide evidence that SAFE ma-QAOA can accelerate convergence to high-quality solutions while reducing the required quantum resources for exact fine-tuning,
offering a resource-efficient route toward expressive ma-QAOA on NISQ hardware.
\end{abstract}

\begin{IEEEkeywords}
Quantum Approximate Optimization Algorithm, multi-angle QAOA, low-weight Pauli propagation, parameter distillation, spin glass, Sherrington-Kirkpatrick model, two-dimensional square lattice
\end{IEEEkeywords}

\section{Introduction}
\label{sec:introduction}
The Quantum Approximate Optimization Algorithm (\qaoa) is a variational framework for solving combinatorial optimization problems on near-term quantum devices \cite{farhi2014quantumapproximateoptimizationalgorithm}.
By alternating a problem-specific cost Hamiltonian with a mixing Hamiltonian, \qaoa\ directly encodes classical objectives and has been widely applied to graph and Ising-type models \cite{crooks2018performancequantumapproximateoptimization,PhysRevX.10.021067,BLEKOS20241}.
In standard \qaoa, all cost terms in a layer share a single angle $\gamma_\ell$, and all mixer terms in the same layer share a single angle $\beta_\ell$.
This shared-angle structure keeps the ansatz compact, but improving expressivity often requires increasing circuit depth.
On NISQ hardware, deeper circuits make each objective or gradient evaluation more demanding and can worsen trainability through plateaus \cite{Uvarov_2021,Wang2021}.
Even low-depth QAOA landscapes can exhibit unfavorable local structure \cite{10.1145/3649153.3649204}.
Deeper circuits also amplify noise sensitivity as gate errors and decoherence accumulate with circuit depth \cite{Wang2021,Weidenfeller2022scalingofquantum,PhysRevResearch.6.013223,Pelofske2024,Lotshaw2022,PAN202211273,Pellow-Jarman2024}.

Multi-angle \qaoa\ (ma-\qaoa) and related variants address this expressivity-depth tradeoff by introducing additional variational parameters at low circuit depths \cite{Herrman2022,Jang_2026,Vijendran_2024}.
This term-wise parameterization can improve expressivity at low depth, making ma-\qaoa\ attractive in low-depth NISQ settings.
However, this flexibility substantially increases the number of trainable angles.
Direct exact optimization must refine a high-dimensional ma-\qaoa\ parameter vector, increasing the cost of objective and gradient evaluations.
In practice, the exact fine-tuning cost of ma-\qaoa\ depends on two quantities: the number of active angles and the number of exact optimization steps needed to reach a high-quality solution.
Existing approaches address this through warm-start schedules, parameter-initialization heuristics, lower-dimensional parameterizations, and surrogate or adaptive optimization \cite{Sack2021quantumannealing,Gaidai2024,brandao2018fixedcontrolparametersquantum,Cheng2024}, but expressive ma-\qaoa\ training can still require many evaluations of the exact energy objective.

We address this problem with Surrogate-Assisted and Fine-tuning Enhanced (SAFE) ma-\qaoa, which combines \lwpp-based surrogate pre-training, parameter distillation, and exact fine-tuning.
\lwpp\ is a scalable classical surrogate that evaluates expectation values without full state-vector simulation, at a cost that grows polynomially rather than exponentially in qubit count \cite{4kyq-n8jb}.
However, we find that \lwpp\ alone does not reach the solution quality achievable by exact optimization, making an exact fine-tuning stage necessary.
In SAFE, we therefore use \lwpp\ to pre-train the full ma-\qaoa\ parameter set, then initialize exact fine-tuning with the resulting parameters.
We study two SAFE variants: with distillation, where negligible parameters are removed after pre-training, and without distillation.

The remainder of the paper is organized as follows. Section~\ref{sec:background} reviews \qaoa, ma-\qaoa, and Pauli propagation.
Section~\ref{sec:experimental-setup} describes the problem families, initialization settings, metrics, and training protocol.
Section~\ref{sec:results} presents the empirical results.
Section~\ref{sec:discussion} discusses their implications.
Section~\ref{sec:conclusion} concludes the paper.

\section{Background}
\label{sec:background}

\subsection{Quantum Approximate Optimization Algorithm (QAOA) and Multi-Angle QAOA}
\label{subsec:qaoa}

\subsubsection{Standard QAOA}
The cost Hamiltonian $H_C$ is problem-dependent and encodes the objective of interest; here we identify it with the target Hamiltonian $\hat{H}$.
The mixing Hamiltonian is chosen as the standard transverse-field mixer,
\begin{equation}
H_C = \hat{H}, \quad H_M = \sum_{i=1}^n X_i,
\label{eq:hamiltonians}
\end{equation}
where $n$ denotes the number of qubits.
The corresponding evolutions are implemented by the unitaries $U_C(\gamma)$ and $U_B(\beta)$:
\begin{equation}
U_C(\gamma) = e^{-i\gamma H_C}, \quad U_B(\beta) = e^{-i\beta H_M}.
\label{eq:unitaries}
\end{equation}

Here, $p$ denotes the circuit depth, i.e., the number of alternating cost and mixer layers.
The resulting depth-$p$ ansatz prepares a trial state $\ket{\psi(\boldsymbol{\gamma},\boldsymbol{\beta})}$ by applying $U_C$ and $U_B$ for $p$ layers to the uniform superposition state $\ket{+}^{\otimes n}$:
\begin{equation}
\ket{\psi(\boldsymbol{\gamma},\boldsymbol{\beta})}
=
U_B(\beta_p)U_C(\gamma_p)\cdots U_B(\beta_1)U_C(\gamma_1)\ket{+}^{\otimes n},
\end{equation}
where $\boldsymbol{\gamma}$ and $\boldsymbol{\beta}$ are variational parameters.
In QAOA training, the circuit is evaluated repeatedly: the expectation value $E(\boldsymbol{\gamma},\boldsymbol{\beta}) = \bra{\psi(\boldsymbol{\gamma},\boldsymbol{\beta})} H_C \ket{\psi(\boldsymbol{\gamma},\boldsymbol{\beta})}$ is estimated for the current parameters,
a classical optimizer updates $\boldsymbol{\gamma}$ and $\boldsymbol{\beta}$, and this loop is repeated until a stopping criterion is met.
This shared parameterization keeps the circuit compact, but constrains expressivity at a given depth.

\subsubsection{Multi-Angle QAOA}
While standard QAOA uses one shared pair $(\gamma_\ell,\beta_\ell)$ per layer, multi-angle QAOA (ma-QAOA) assigns separate trainable angles to each term in the cost and mixer layers \cite{Herrman2022,Gaidai2024,9996634,Jang_2026}.
This term-wise parameterization increases the expressivity of the ansatz at low circuit depths. Writing the cost Hamiltonian as a weighted Pauli decomposition $H_C=\sum_{\alpha=1}^{M_C} c_{\alpha} P_{\alpha}$, where $P_{\alpha}$ is a Pauli operator and $c_{\alpha}$ is its coefficient, the ma-QAOA layer unitaries are generalized as:
\begin{equation}
U_C^{(\ell)} = \prod_{\alpha=1}^{M_C} e^{-i \gamma_{\alpha,\ell} P_{\alpha}}, \quad U_B^{(\ell)} = \prod_{i=1}^{n} e^{-i \beta_{i,\ell} X_i}.
\end{equation}
Here, $\gamma_{\alpha,\ell}$ denotes the effective term-wise cost rotation angle associated with Pauli term $P_{\alpha}$ at layer $\ell$ under a fixed product ordering, and $\beta_{i,\ell}$ is the angle for mixer term $X_i$ at layer $\ell$. However, this flexibility also changes the training bottleneck: instead of optimizing a small number of shared angles, ma-QAOA exposes a $p(M_C+n)$-dimensional parameter space whose objective values and gradients must be estimated repeatedly throughout the classical optimization loop.
On hardware, each estimate consumes circuit executions and shots, and gradient estimation multiplies this cost across many independently tuned angles, making evaluation efficiency a central concern even at modest circuit depths.

Prior work has addressed this from three directions.
Warm-start schedules and parameter-transfer heuristics reduce the number of exact optimization steps by providing informed starting points \cite{Sack2021quantumannealing,PhysRevX.10.021067,apte2025iterativeinterpolationschedulesquantum,10.1145/3676536.3697128,9605328,patel2026improvingefficiencyqaoausing}.
Fixed or lower-dimensional parameterizations can reduce the number of angles requiring instance-specific optimization \cite{brandao2018fixedcontrolparametersquantum,PhysRevA.104.L010401,sakai2025transferringlinearlyfixedqaoa}.
Surrogate and adaptive optimization approaches learn approximate models from limited quantum evaluations to reduce per-step evaluation cost \cite{tibaldi2023bayesianoptimizationqaoa,self_variational_2021,PhysRevA.107.032415,Smith_2023,Cheng2024}.
In the ma-\qaoa\ setting, these considerations motivate classical surrogates that guide the initial training phase and reserve expensive exact or hardware evaluations for a targeted fine-tuning stage \cite{PhysRevA.99.032331,Wierichs2022generalparameter,Kubler2020adaptiveoptimizer,arrasmith2020operatorsamplingshotfrugaloptimization,24gg-7p8z}.

\subsection{Pauli Propagation}
\label{subsec:pauli-propagation}
Pauli propagation is a computational framework for evaluating expectation values that operates directly in the Heisenberg picture \cite{PhysRevA.99.062337,rudolph2025paulipropagationcomputationalframework,angrisani2025simulatingquantumcircuitsarbitrary,Fontana2025}.
Unlike traditional state-vector emulation, which tracks the evolution of the full quantum state $\ket{\psi(\boldsymbol{\theta})} = U(\boldsymbol{\theta})\ket{\phi_0}$ in the Schr\"odinger picture,
Pauli propagation keeps the input state fixed and evolves the measured observable $\hat{\mathrm{O}}$ backward through the circuit.

More explicitly, if $\ket{\psi(\boldsymbol{\theta})} = U(\boldsymbol{\theta})\ket{\phi_0}$, then the expectation value can be written as:
\begin{equation}
\langle \hat{\mathrm{O}} \rangle
=
\bra{\phi_0} U^\dagger(\boldsymbol{\theta}) \hat{\mathrm{O}} U(\boldsymbol{\theta}) \ket{\phi_0}
.
\end{equation}
This viewpoint is natural for variational optimization because the objective depends only on $\hat{\mathrm{O}}$, not on the full output state.
Since the $n$-qubit Pauli operators form a complete orthogonal basis for the space of Hermitian operators, $\hat{\mathrm{O}}$ can be decomposed as a weighted sum of tensor-product Pauli operators $P_{\alpha} \in \{I,X,Y,Z\}^{\otimes n}$:
\begin{equation}
\hat{\mathrm{O}} = \sum_{\alpha} c_{\alpha} P_{\alpha}.
\end{equation}
The evolution of $\hat{\mathrm{O}}$ under a parameterized Pauli rotation $R_G(\theta) = e^{-i\theta G/2}$ is then computed via conjugation.
For each Pauli string $P$, the update rule is determined by the commutation relation between $P$ and the generator $G$:
\begin{equation}
R_G^\dagger(\theta) P R_G(\theta)
=
\left\{
\begin{array}{@{}l@{\quad}l@{}}
P, & [P,G]=0,\\
\cos\theta\,P + \sin\theta\,P', & \{P,G\}=0,
\end{array}
\right.
\end{equation}
where $P' = -iPG$ is another Pauli string. Each anticommuting gate splits one tracked term into two, so the total number of tracked Pauli strings can grow exponentially with circuit depth.

To maintain computational tractability, low-weight Pauli Propagation (LWPP) introduced a systematic truncation strategy \cite{4kyq-n8jb}.
In this approach, any Pauli string whose weight (the number of qubits on which it carries a non-identity operator) exceeds a predefined threshold $w_{\max}$ is discarded during the propagation process.
Fig.~\ref{fig:pps-branching-tree} illustrates this truncation, where higher-weight operators are discarded to preserve classical tractability.
By controlling $w_{\max}$, LWPP acts as a scalable surrogate for calculating expectation values, which we employ in this work to reduce the evaluation cost of ma-QAOA optimization.

\setcounter{figure}{0}
\begin{figure}[t]
\centering
\includegraphics[width=\columnwidth]{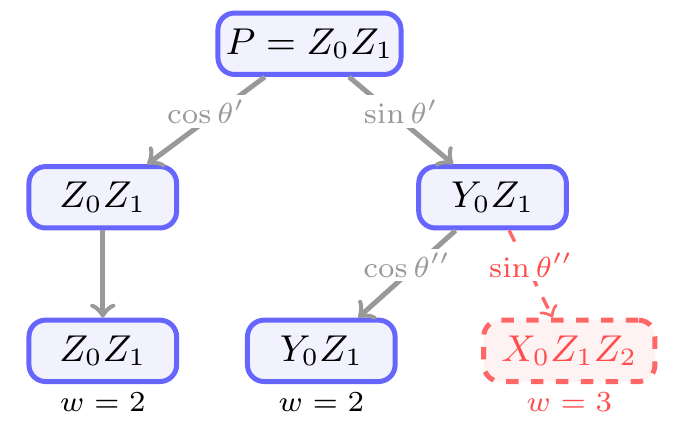}
\caption{Pauli propagation example for the tracked Pauli word $P=Z_0Z_1$ under successive updates by $R_{X_0}(\theta')$ and $R_{Z_0Z_2}(\theta'')$. The example shows the lower-weight branches retained by LWPP and the higher-weight branch discarded by the $w_{\max}=2$ truncation threshold.}
\label{fig:pps-branching-tree}
\end{figure}

\setcounter{figure}{1}
\begin{figure*}[t!]
\centering
\includegraphics[width=\textwidth]{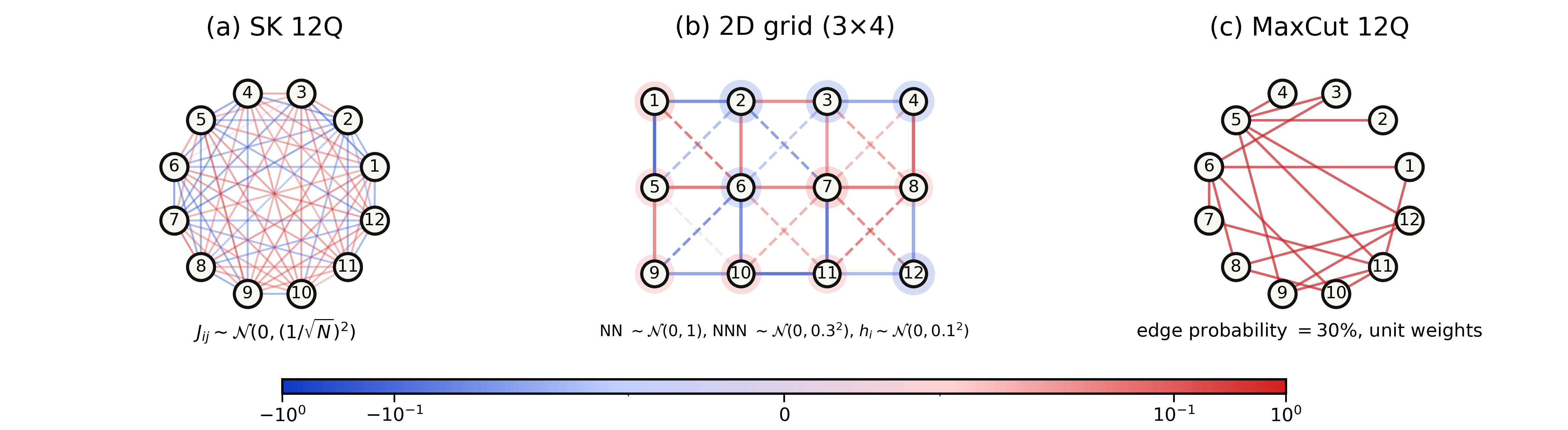}
\caption{Problem families used in the experimental setup: a fully connected SK instance with 12 qubits, a two-dimensional square-lattice spin glass (abbreviated as 2D grid in figures and tables) on a $3\times 4$ lattice, and a 12-qubit Max-Cut instance with 30\% edge probability and unit weights.}
\label{fig:problem-family-schematic-1x3}
\end{figure*}

\subsection{Scalability of the LWPP Surrogate}
\label{subsec:lwpp-scalability}
The scalability of LWPP follows from the bounded number of Pauli strings retained after truncation~\cite{4kyq-n8jb}.
Without truncation, Pauli propagation can generate Pauli strings over the full $n$-qubit Pauli basis, whose size is $4^n$.
In contrast, LWPP retains only Pauli strings with weight at most $w_{\max}$.
The number of such strings is bounded:
\begin{equation}
N_{\mathrm{LWPP}}(n,w_{\max})
\leq
\sum_{r=0}^{w_{\max}} \binom{n}{r}3^r,
\end{equation}
because one first chooses the $r$ non-identity qubit locations and then assigns one of $\{X,Y,Z\}$ to each selected location.
For fixed $w_{\max}$, this bound scales as
\begin{equation}
N_{\mathrm{LWPP}}(n,w_{\max}) = O(n^{w_{\max}}),
\end{equation}
so the number of propagated Pauli strings retained by the truncated surrogate grows polynomially with the number of qubits rather than exponentially with the size of the full Pauli basis.
Because GPU execution time varies across hardware, we report the number of propagated Pauli strings that must be processed during LWPP.
Empirically, the counts of propagated Pauli strings stay well within this bound: for the fully connected SK cost Hamiltonian at $w_{\max}=4$, the counts are $2.3\times10^{4}$, $8.1\times10^{4}$, and $2.1\times10^{5}$ at $n=12$, $16$, and $20$, respectively, corresponding to about half of the bound at each size (and roughly $45\%$ at $w_{\max}=3$). These counts are many orders of magnitude below the untruncated $4^{20}\approx10^{12}$ basis at $n=20$; the 2D-grid and Max-Cut counts constitute an even smaller fraction of the corresponding bound.
This is how LWPP enables scalable surrogate pre-training: the expensive early optimization stage can be efficiently performed without full state-vector simulation, while exact evaluation is reserved for the later fine-tuning stage.

\section{Experimental Setup}
\label{sec:experimental-setup}

\subsection{Problem Instances Construction}
\label{subsec:problem-instances}
We evaluate SAFE on three problem families spanning qualitatively different interaction structures --- a fully connected model, a sparse geometric lattice, and a random sparse graph --- to assess how broadly the workflow generalizes.
The problem instances evaluated in this work are encoded as Ising-type cost Hamiltonians \cite{10.3389/fphy.2014.00005,Mohseni2022,PhysRevX.5.031026,Fan2023}:
\begin{equation}
H_C = \sum_{i} h_i Z_i + \sum_{(i,j) \in E} J_{ij} Z_i Z_j,
\label{eq:general-ising}
\end{equation}
where $h_i$ represents the local longitudinal fields and $J_{ij}$ denotes the coupling strengths between spins on a graph $G=(V,E)$.

We evaluate our framework across problem scales ranging from 12 to 20 qubits, in increments of 4 qubits for the SK and Max-Cut families. For the two-dimensional square-lattice spin glass, abbreviated as 2D grid in figures and tables, we use the $3\times 4$, $4\times 4$, and $4\times 5$ lattices. This size range reflects the practical limits of exact state-vector emulation and brute-force ground-state search, both of which become costly beyond 30 qubits; the LWPP pre-training stage is not subject to this constraint, as the number of propagated Pauli strings grows polynomially with the qubit count for fixed $w_{\max}$. For each problem size across all families, we apply five random instances to ensure the generality of our findings.
For these Ising-type instances, the ma-\qaoa\ cost angles are assigned to the active $Z_i$ and $Z_iZ_j$ terms, and the mixer angles are assigned to the $n$ single-qubit $X_i$ terms in each layer.

\begin{itemize}
    \item \emph{Sherrington-Kirkpatrick (SK) Model}: A prototypical fully connected spin glass \cite{Sherrington2025,Panchenko2012}. We set $h_i = 0$ and draw the couplings $J_{ij}$ from a Gaussian distribution $\mathcal{N}(0, 1/n)$. This dense, all-to-all connectivity provides a rigorous test for navigating highly frustrated energy landscapes \cite{PhysRevX.5.031040,Farhi2022quantumapproximate,https://doi.org/10.4230/lipics.tqc.2022.7,boulebnane2025evidencequantumapproximateoptimization,montanari2019optimizationsherringtonkirkpatrickhamiltonian}.
    \item \emph{Two-Dimensional Square-Lattice Spin Glass}: A sparse square-lattice model \cite{PhysRevB.65.134404,PhysRevE.78.056705,chatterjee2026spinglassphasezero} with nearest- and next-nearest-neighbor couplings \cite{PhysRevE.80.051117,Fytas_2008}. We sample nearest-neighbor couplings from $\mathcal{N}(0,1)$, next-nearest-neighbor couplings from $\mathcal{N}(0,0.3^2)$, and local fields $h_i$ from $\mathcal{N}(0,0.1^2)$.
    \item \emph{Max-Cut Problem}: A combinatorial optimization benchmark with sparse, unweighted interactions, mapped to the Ising framework by setting $h_i = 0$ and $J_{ij} = 1$ for all $(i,j) \in E$. We generate instances using Erd\H{o}s--R\'{e}nyi random graphs with a 30\% edge probability.
\end{itemize}

\subsection{Initialization Settings}
\label{subsec:initialization}
To assess the robustness of our training framework, we apply 11 distinct initialization seeds to each instance.
Among these, only the \emph{QAOA Relax} strategy uses a warm start derived from a discretized quantum annealing schedule (\tqa) \cite{Sack2021quantumannealing,PhysRevX.10.021067,hxv2-sbr7,nzongani2026scalingqaoatransferringoptimal}.
\begin{itemize}
    \item \emph{Random Initialization (5 seeds)}: Five independent sets of random initial parameters $(\boldsymbol{\gamma}, \boldsymbol{\beta})$ sampled uniformly from $[-0.5, 0.5]$.
    \item \emph{Constant Initialization (5 seeds)}: Five different uniform starts where all $\gamma$ angles are set to a positive constant and all $\beta$ angles to the corresponding negative constant, with magnitudes $0.01$, $0.05$, $0.1$, $0.2$, and $0.4$.
    \item \emph{QAOA Relax (1 seed)}: This strategy starts from a trotterized quantum annealing schedule
    \begin{equation}
    H_{\mathrm{ann}}(s) = (1-s)H_M + sH_C,
    \end{equation}
    using the cost and mixing Hamiltonians defined in \eqref{eq:hamiltonians}. With $s \in [0,1]$, we discretize the schedule by setting $s_\ell = \ell/p$ and initialize, for $\ell=1,\dots,p$,
    \begin{equation}
    \gamma_\ell = \frac{\ell}{p}\Delta t, \quad
    \beta_\ell = -\left(1-\frac{\ell}{p}\right)\Delta t,
    \end{equation}
    with $\Delta t = 0.5$. The resulting parameters are then relaxed with a short classical optimization pass before entering the main training loop.
\end{itemize}

\subsection{Training Protocol and Parameter Distillation}
\label{subsec:training-protocol}

The SAFE training pipeline consists of three sequential stages: LWPP surrogate pre-training, optional parameter distillation, and exact fine-tuning (Fig.~\ref{fig:safe-pipeline}).
We compare two SAFE variants: with distillation, where negligible parameters are
removed after pre-training, and without distillation.
Performance across these settings is evaluated using the metrics defined in Section~\ref{subsec:metric}.

\setcounter{figure}{2}
\begin{figure}[!t]
\centering
\includegraphics[width=\columnwidth]{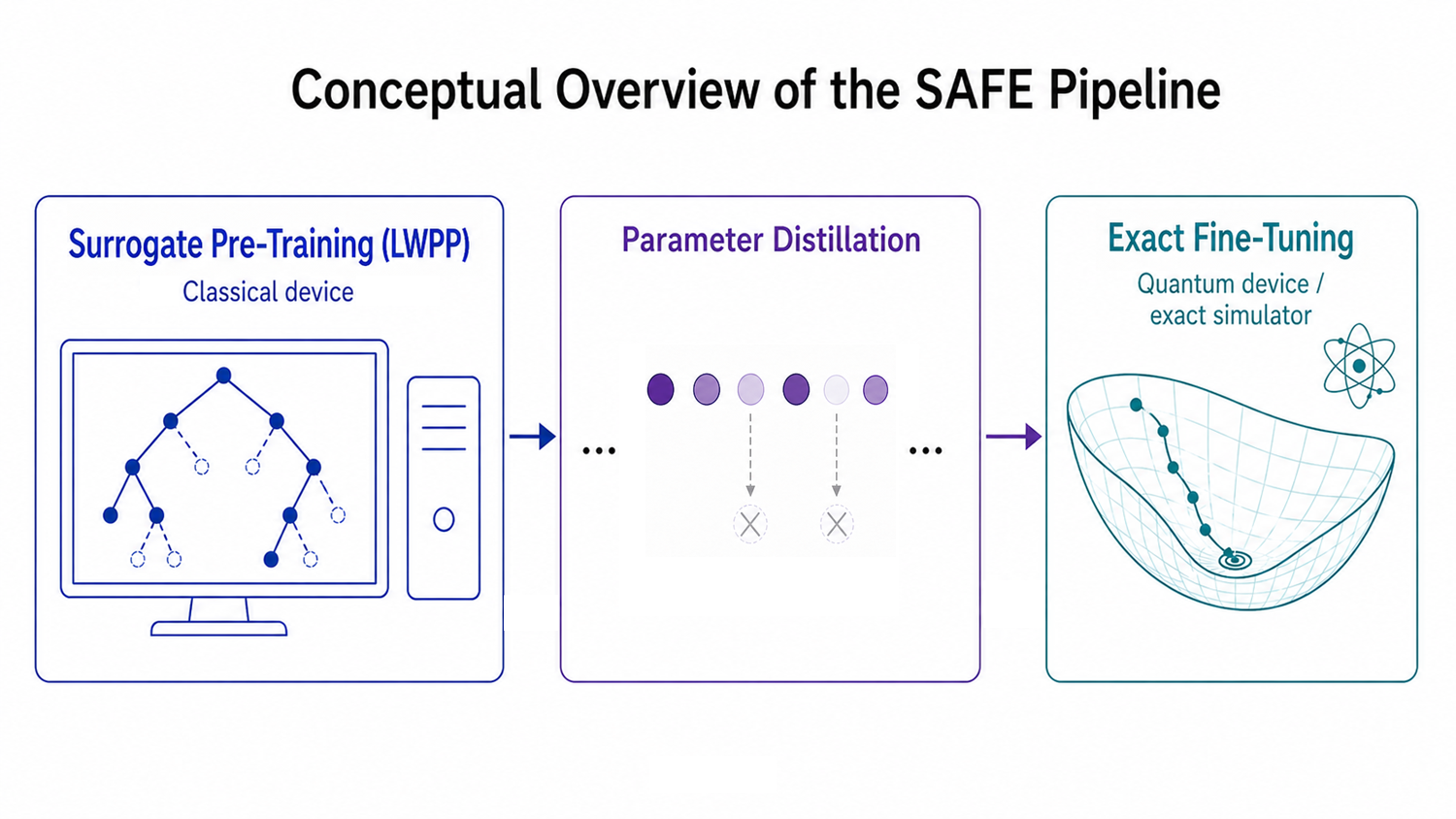}
\caption{Conceptual overview of the SAFE pipeline: LWPP-based surrogate pre-training on classical hardware, followed by parameter distillation to identify a compact subspace, and fine-tuning on quantum hardware or exact emulators.}
\label{fig:safe-pipeline}
\end{figure}

\subsubsection{LWPP Pre-Training}
The pre-training phase consists of 500 steps using LWPP as a surrogate model. This computational budget is chosen to allow the surrogate objective to reach approximate convergence while remaining substantially cheaper than exact evaluation; in practice, the surrogate loss plateau is typically reached before this limit. We restrict the maximum Pauli weight to $w_{\max}\in\{3,4\}$, spanning lighter and moderate truncation levels to assess whether heavier truncation is necessary for denser problem families. The pre-training objective is evaluated on a truncated operator representation before the resulting parameters are used for exact fine-tuning.

\subsubsection{Parameter Distillation}
We compare three distillation thresholds: $0$, which keeps all parameters, and $0.01$ or $0.3$, which remove parameters below the corresponding absolute value threshold.

\subsubsection{Exact Fine-Tuning}
In this work, exact optimization refers to noiseless classical optimization of the ma-\qaoa\ energy objective. The exact-only baseline starts directly from the original parameter initialization, whereas SAFE starts the same optimization from parameters obtained through LWPP pre-training.
After LWPP pre-training, we pass either the full parameter set or the parameters retained after distillation to an exact evaluator. We then perform 100 additional fine-tuning steps on the exact objective. This budget is chosen to be long enough to observe convergence behavior across most initialization settings, while reflecting a practical constraint on exact optimization cost.

For each of the eleven initialization settings, exact-only uses the same 100-step exact budget but skips LWPP pre-training and distillation. This ensures the comparison isolates the effect of surrogate-guided initialization rather than the amount of exact computation available to each method.

\subsubsection{Implementation Details}
All classical optimization stages use the Adam optimizer with a learning rate of $0.02$ and default betas $(0.9, 0.999)$.
All experiments are implemented in Python~3.11 with PyTorch~2.5.1 for gradient computation and CUDA~12 for GPU acceleration. The LWPP surrogate is built on our custom Pauli propagation library \cite{normaq_pp_gpu}, publicly available at \href{https://github.com/Norma-Q/PADO-Pauli}{\texttt{https://github.com/Norma-Q/PADO-Pauli}}.

\subsection{Performance Metrics: Approximation Ratio, First-Hit Step, and Fine-Tuning Cost Ballpark Estimate}
\label{subsec:metric}
To capture complementary aspects of SAFE's performance --- final solution quality, convergence speed, and estimated hardware cost --- we define the following three metrics.
\begin{itemize}
    \item \emph{Approximation ratio}: We define a normalized approximation ratio $\alpha$ that maps the expectation value to a range between 0 and 1. For a given problem instance, let $E_{\min}$ and $E_{\max}$ denote the minimum (ground state) and maximum energy levels, respectively, obtained through exhaustive brute-force search. The approximation ratio for the current variational parameters $\boldsymbol{\theta}$ is then defined as:
\begin{equation}
\alpha = \frac{E_{\max} - E(\boldsymbol{\theta})}{E_{\max} - E_{\min}},
\label{eq:approx-ratio}
\end{equation}
where $E(\boldsymbol{\theta}) = \bra{\psi(\boldsymbol{\gamma}, \boldsymbol{\beta})} H_C \ket{\psi(\boldsymbol{\gamma}, \boldsymbol{\beta})}$ denotes the expectation value within the ma-QAOA framework, with $\boldsymbol{\theta} = (\boldsymbol{\gamma}, \boldsymbol{\beta})$ representing the set of variational parameters.

By this definition, $\alpha = 1$ corresponds to the exact discovery of the global minimum, while $\alpha = 0$ represents the worst possible state. This normalization is essential because random spin-glass instances can have vastly different energy scales depending on their coupling distributions and connectivity. Using $\alpha$ allows us to provide a consistent comparison of convergence efficiency and approximation ratio performance, independent of the intrinsic energy bounds of the specific instance.

    \item \emph{First-hit step}: To compare convergence speed during exact fine-tuning, we also define the first-hit step $\tau_{0.99}$.
For an approximation ratio trajectory $\alpha(t)$ for a time step $t$ and the best value $\alpha_{\mathrm{best}}$ reached by the same method during exact fine-tuning, $\tau_{0.99}$ is defined as:
\[
\tau_{0.99}=\min\{t : \alpha(t)\ge 0.99 \alpha_{\mathrm{best}}\}.
\]
This metric measures how quickly each method enters its own near-optimal regime. Measuring against each method's own best rather than a common target ensures that convergence speed is evaluated independently of final solution quality. It is therefore used to compare fine-tuning speed rather than final approximation ratio performance.

    \item \emph{Fine-tuning cost ballpark estimate}: To account for the parameter dimension of the exact fine-tuning stage, we also define a ballpark estimate of fine-tuning cost,
\[
C_{\mathrm{ballpark}}=N_{\mathrm{active}}\tau_{0.99},
\]
where $N_{\mathrm{active}}$ is the number of trainable parameters used during the exact optimization phase. For exact-only and SAFE without distillation, $N_{\mathrm{active}}$ corresponds to the full ma-\qaoa\ parameter count, whereas for SAFE with distillation it represents the reduced parameter count after distillation. We focus this estimate on the exact optimization phase because LWPP pre-training is a classical surrogate stage whose cost is controlled by operator truncation, while exact fine-tuning requires repeated evaluations of the exact energy objective.
This ballpark estimate should not be interpreted as an exact hardware runtime or a literal QPU workload, since the actual execution cost depends on shot allocation, measurement grouping, gradient-estimation rules, and device-specific overheads.
Instead, it gives a rough sense of how fine-tuning cost scales with both the number of active trainable parameters and the first-hit step.
\end{itemize}

\begin{figure*}[t]
\centering
\includegraphics[width=0.98\textwidth]{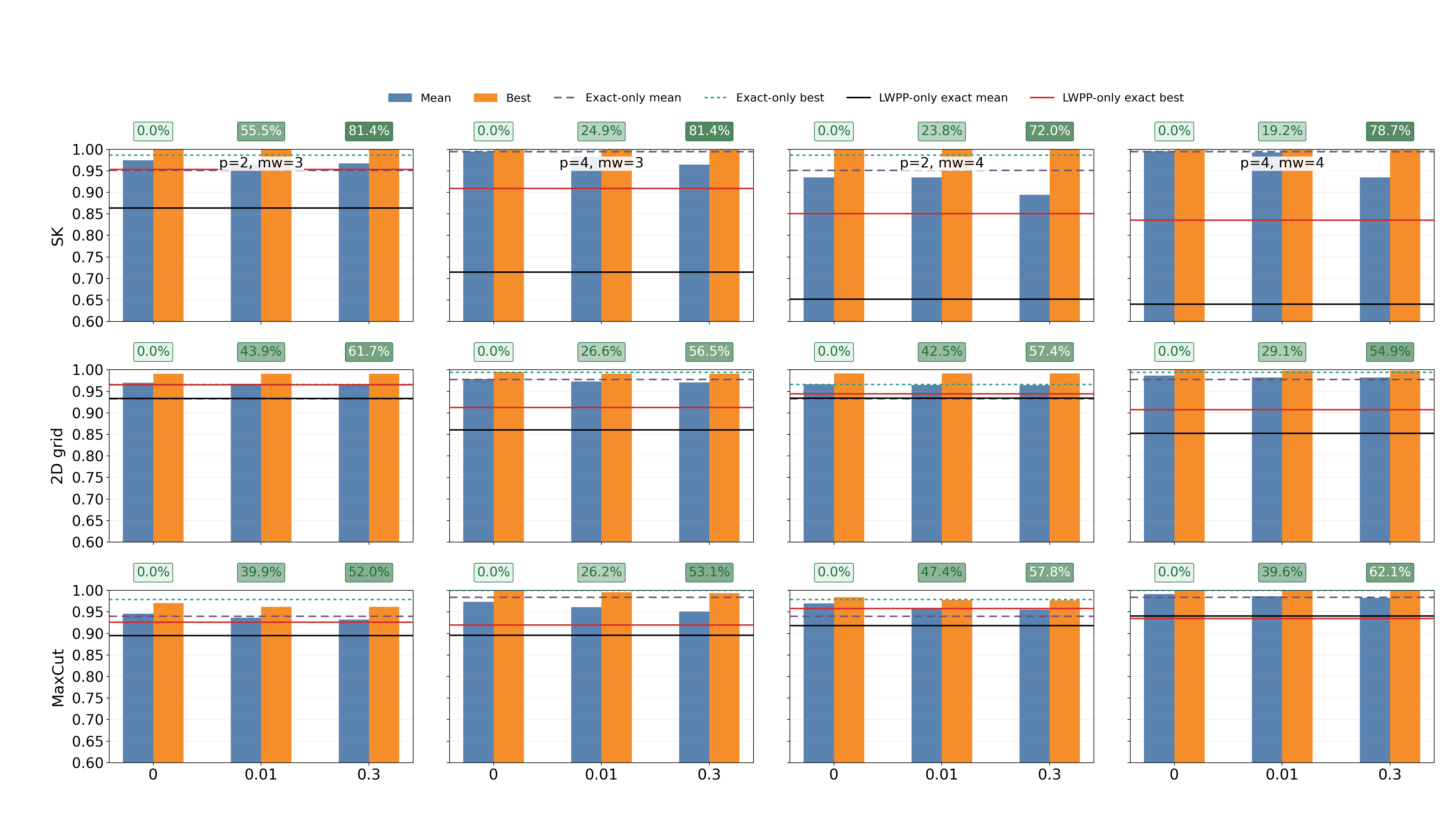}
\caption{Overview of final approximation ratio performance after exact fine-tuning initialized with parameters obtained from LWPP pre-training under different distillation thresholds. Rows correspond to SK, 2D grid, and Max-Cut, and columns compare depths $p=2,4$ with LWPP truncation levels $w_{\max}=3,4$. For each threshold, the bars report the mean and best final approximation ratios, the horizontal lines indicate the exact-only and LWPP-pre-training-only reference values, and the green labels show the average parameter reduction.}
\label{fig:results-overview}
\end{figure*}

\begin{figure*}[!t]
\centering
\includegraphics[width=0.98\textwidth]{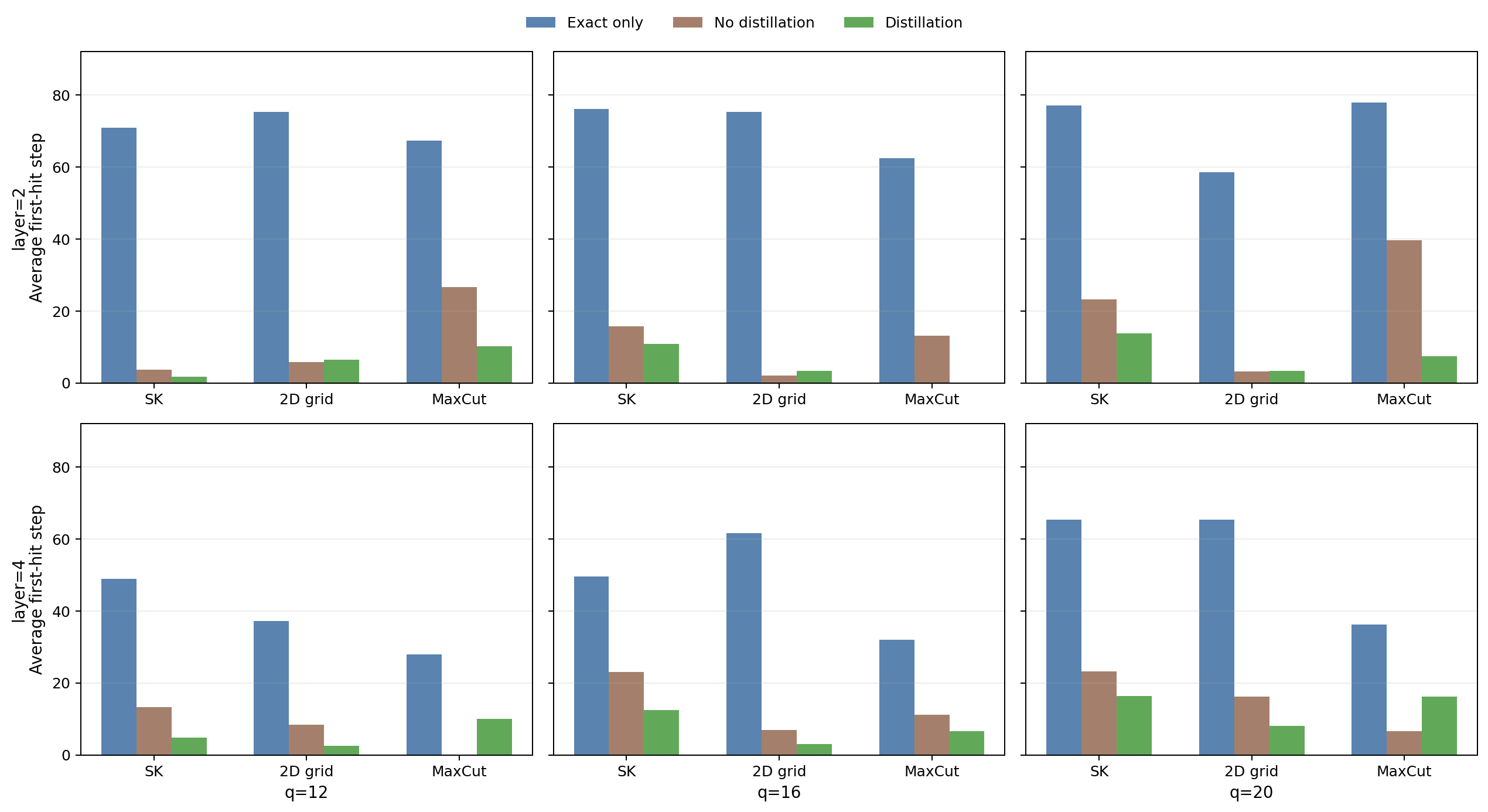}
\caption{Average first-hit step required to reach $99\%$ of each method's own best normalized performance during exact fine-tuning. Rows compare depths $p=2$ and $p=4$, columns correspond to system sizes $n=12,16,20$, and each panel reports SK, 2D grid, and Max-Cut results for exact-only, SAFE without distillation, and SAFE with distillation under $w_{\max}=4$ and a distillation threshold of $0.3$. Lower values indicate faster convergence to the near-optimal regime.}
\label{fig:first_hit_step}
\end{figure*}

\section{Results}
\label{sec:results}
\subsection{Approximation Ratio and Parameter Reduction Analysis}
\label{subsec:distillation-overview}
Distillation maintains strong approximation ratio performance while removing many ma-\qaoa\ parameters.
Fig.~\ref{fig:results-overview} summarizes the 20-qubit distillation results for the SK, 2D grid, and Max-Cut families.
Following the convention defined in Section~\ref{subsec:training-protocol}, SAFE without distillation and SAFE with distillation differ only by whether parameter distillation is applied before exact fine-tuning.
We compare two depths, $p\in\{2,4\}$, two LWPP truncation levels, $w_{\max}\in\{3,4\}$, and three distillation thresholds, $0$, $0.01$, and $0.3$ (in x-axis order from left to right in each panel in Fig.~\ref{fig:results-overview}),
where $0$ denotes no distillation, while $0.01$ and $0.3$ remove parameters whose absolute values are below their respective values.
The blue and orange bars report the mean and best final approximation ratios across the eleven initialization seeds after fine-tuning,
and the green labels show the average parameter-reduction fraction.

The main observation is that the SAFE workflow with distillation can remove a large proportion of the ma-\qaoa\ parameters while preserving high final approximation ratio performance.
At the $0.3$ threshold, the parameter-reduction fraction reaches the 50--80\% range across the 20-qubit settings (average 64.3\%),
yet the best-case approximation ratios remain close to the threshold-$0$ SAFE setting.
This robustness of the best case is the key result of Fig.~\ref{fig:results-overview}: even after aggressive distillation, the pipeline can still identify a compact parameterization
that fine-tunes to a high final approximation ratio.

The comparison with exact-only further shows that distilled SAFE can still achieve strong final approximation ratio performance after parameter removal.
In Fig.~\ref{fig:results-overview}, the orange bars represent the best final approximation ratios after exact fine-tuning, while the dashed horizontal lines mark the corresponding exact-only references.
Even at the aggressive threshold $0.3$, where the green labels show substantial parameter reduction, the best distilled SAFE bars remain close to the exact-only best lines in most panels and reach or exceed them in several SK and 2D grid settings.
This pattern is most visible in the best-case bars. It supports the interpretation that LWPP pre-training provides an effective starting point for exact fine-tuning, and that distillation further preserves access to high-performing parameter regimes while removing low-magnitude angles.

However, LWPP pre-training alone is not sufficient to reach the final approximation ratio achieved by SAFE after exact fine-tuning.
This point is visible in Fig.~\ref{fig:results-overview}, where the LWPP-only mean and best lines remain well below the final SAFE bars after exact fine-tuning in many panels.
Therefore, LWPP should not be interpreted as a replacement for classical or QPU-based optimization. Its role in SAFE is to guide parameters to the near-optimal regime while the fine-tuning stage is necessary to achieve high approximation ratios. Further analysis will be discussed in Section~\ref{subsec:param-similarity}.

The approximation ratio results indicate that the best LWPP truncation weight differs across problem families and instances.
For the 2D grid and SK families, the bars at $w_{\max}=3$ already remain near the exact-only references across both depths, indicating that lighter truncation is sufficient for them.
For Max-Cut, the two truncation levels are closer overall, although $w_{\max}=4$ gives the more stable visual pattern in the distilled settings.
These results support treating $w_{\max}$ as a problem-dependent hyperparameter rather than a universal constant.

\subsection{First-Hit Step and Fine-Tuning Cost Ballpark Estimate Analysis}
\label{subsec:first-hit-step}

After showing that SAFE with distillation can remove a substantial fraction of trainable parameters while preserving strong final approximation ratio performance,
we next ask whether this compact parameterization also reaches the near-optimal regime more rapidly.
Fig.~\ref{fig:first_hit_step} therefore focuses on convergence speed with the first-hit step $\tau_{0.99}$ defined in Section~\ref{subsec:metric}.
Here, we fix $w_{\max}=4$ and the distillation threshold at $0.3$, and compare three settings: exact-only, SAFE without distillation, and SAFE with distillation.
The y-axis reports the average fine-tuning step, averaged over the corresponding problem instances and initialization settings, at which each method first reaches $99\%$ of its own best normalized performance,
so the comparison targets convergence speed rather than final approximation ratio performance itself.

LWPP pre-training substantially reduces the first-hit step during exact fine-tuning.
Across SK, 2D grid, and Max-Cut, Fig.~\ref{fig:first_hit_step} shows that the exact-only bars are generally much higher, whereas SAFE without distillation already reaches the near-optimal regime in substantially fewer steps.
For smaller settings, this acceleration can occur because LWPP pre-training already captures a high-quality exact-energy configuration before fine-tuning. For example, Max-Cut at $(n=12,\;p=4)$ reaches the near-optimal regime at step zero without distillation, corresponding to the step-0 approximation ratio of $1.000$ in Table~\ref{tab:init-to-final-ratio}.
More importantly, the step reduction is not limited to such easy cases.
It remains visible at $n=20$ and persists at the deeper $p=4$ setting, where the gap between the LWPP pre-trained step-0 energy and the final fine-tuned energy can still be large.
Thus, even when the gap between the LWPP-only and exact fine-tuning energies is large, LWPP pre-training turns out to be useful.

\begin{table*}[t]
\centering
\caption{Best-case approximation-ratio progression during exact fine-tuning for $w_{\max}=4$. Each entry reports the step-0 value, the best value reached during exact fine-tuning, and the gain.}
\label{tab:init-to-final-ratio}
\scriptsize
\setlength{\tabcolsep}{5pt}
\renewcommand{\arraystretch}{1.12}
\resizebox{\textwidth}{!}{%
\begin{tabular}{ll|ccc|ccc}
\hline
Family & Method & \multicolumn{3}{c|}{$p=2$} & \multicolumn{3}{c}{$p=4$} \\
\cline{3-8}
 &  & $n=12$ & $n=16$ & $n=20$ & $n=12$ & $n=16$ & $n=20$ \\
\hline
SK & SAFE without distillation & 0.932$\rightarrow$1.000 (+0.068) & 0.850$\rightarrow$0.999 (+0.149) & 0.645$\rightarrow$1.000 (+0.355) & 0.755$\rightarrow$1.000 (+0.245) & 0.634$\rightarrow$1.000 (+0.366) & 0.565$\rightarrow$1.000 (+0.435) \\
SK & SAFE with distillation & 0.968$\rightarrow$1.000 (+0.032) & 0.911$\rightarrow$0.998 (+0.087) & 0.851$\rightarrow$1.000 (+0.149) & 0.946$\rightarrow$1.000 (+0.054) & 0.872$\rightarrow$1.000 (+0.128) & 0.836$\rightarrow$1.000 (+0.164) \\
\hline
2D grid & SAFE without distillation & 0.931$\rightarrow$0.997 (+0.066) & 0.963$\rightarrow$0.991 (+0.028) & 0.939$\rightarrow$0.991 (+0.053) & 0.880$\rightarrow$1.000 (+0.120) & 0.862$\rightarrow$1.000 (+0.138) & 0.850$\rightarrow$1.000 (+0.150) \\
2D grid & SAFE with distillation & 0.924$\rightarrow$0.997 (+0.072) & 0.968$\rightarrow$0.991 (+0.023) & 0.945$\rightarrow$0.991 (+0.047) & 0.973$\rightarrow$1.000 (+0.027) & 0.976$\rightarrow$1.000 (+0.024) & 0.908$\rightarrow$0.998 (+0.090) \\
\hline
Max-Cut & SAFE without distillation & 0.965$\rightarrow$0.993 (+0.028) & 0.960$\rightarrow$0.985 (+0.025) & 0.940$\rightarrow$0.984 (+0.045) & 1.000$\rightarrow$1.000 (+0.000) & 0.948$\rightarrow$1.000 (+0.052) & 0.959$\rightarrow$1.000 (+0.041) \\
Max-Cut & SAFE with distillation & 0.970$\rightarrow$0.985 (+0.016) & 0.977$\rightarrow$0.977 (+0.001) & 0.958$\rightarrow$0.977 (+0.019) & 0.942$\rightarrow$1.000 (+0.058) & 0.978$\rightarrow$0.996 (+0.019) & 0.935$\rightarrow$1.000 (+0.065) \\
\hline
\end{tabular}
}
\end{table*}

Distillation further accelerates this trend in most settings while often improving the starting point obtained from LWPP pre-training.
Among the selected best-case runs reported in Table~\ref{tab:init-to-final-ratio}, SAFE with distillation frequently begins exact fine-tuning from a higher approximation ratio, equivalently a lower exact energy, than SAFE without distillation.
This improvement is especially pronounced for SK: at $n=20$, the step-0 approximation ratio increases from $0.645$ to $0.851$ for $p=2$ and from $0.565$ to $0.836$ for $p=4$ after distillation.
The same mechanism helps explain the broad first-hit step reduction in Fig.~\ref{fig:first_hit_step}: the distilled cases often start closer to the near-optimal regime and then optimize in a smaller parameter space.
This behavior remains clear even in the large cases when $n=20$ or $p=4$, although the effect is not strictly universal.
Moreover, Max-Cut at $(n=16,p=2)$ is a notable case, where SAFE with distillation has first-hit step zero, so the LWPP-pre-trained configuration formed by the parameters retained after distillation is already in the near-optimal regime before exact fine-tuning.
Most settings, however, still require some exact refinement after initialization with parameters obtained from LWPP pre-training, which is why the exact fine-tuning stage remains a necessary part of SAFE rather than being replaced by LWPP alone.

\begin{table*}[!t]
\centering
\caption{Fine-tuning cost ballpark estimates for $w_{\max}=4$ and a distillation threshold of $0.3$. Parentheses report the exact-only-to-SAFE cost ratio, $C_{\mathrm{ballpark}}^{\mathrm{exact}}/C_{\mathrm{ballpark}}^{\mathrm{SAFE}}$. A value of $0.0$ indicates that the LWPP-pre-trained parameters already attain the optimal solution before exact fine-tuning, making the ratio undefined (--).}
\label{tab:ballpark-cost-by-setting}
\scriptsize
\setlength{\tabcolsep}{5pt}
\renewcommand{\arraystretch}{1.15}
\resizebox{\textwidth}{!}{%
\begin{tabular}{ll|ccc|ccc}
\hline
Family & Method & \multicolumn{3}{c|}{$p=2$} & \multicolumn{3}{c}{$p=4$} \\
\cline{3-8}
 &  & $n=12$ & $n=16$ & $n=20$ & $n=12$ & $n=16$ & $n=20$ \\
\hline
SK & Exact-only & 11076.0 & 20726.4 & 32424.0 & 15288.0 & 26982.4 & 54936.0 \\
SK & SAFE without distillation & 561.6 (19.7x) & 4297.6 (4.8x) & 9744.0 (3.3x) & 4118.4 (3.7x) & 12512.0 (2.2x) & 19488.0 (2.8x) \\
SK & SAFE with distillation & 102.8 (107.7x) & 888.6 (23.3x) & 1780.0 (18.2x) & 505.8 (30.2x) & 1844.2 (14.6x) & 3225.0 (17.0x) \\
\hline
2D grid & Exact-only & 7992.4 & 11159.2 & 11134.0 & 7886.4 & 18233.6 & 24852.0 \\
2D grid & SAFE without distillation & 614.8 (13.0x) & 296.0 (37.7x) & 608.0 (18.3x) & 1780.8 (4.4x) & 2072.0 (8.8x) & 6156.0 (4.0x) \\
2D grid & SAFE with distillation & 325.6 (24.5x) & 246.4 (45.3x) & 318.2 (35.0x) & 261.6 (30.1x) & 406.6 (44.8x) & 1251.6 (19.9x) \\
\hline
Max-Cut & Exact-only & 4194.8 & 5964.0 & 12681.6 & 3316.8 & 6540.8 & 11161.6 \\
Max-Cut & SAFE without distillation & 1859.6 (2.3x) & 1267.2 (4.7x) & 6459.6 (2.0x) & 0.0 (--) & 2243.2 (2.9x) & 2027.2 (5.5x) \\
Max-Cut & SAFE with distillation & 428.4 (9.8x) & 0.0 (--) & 533.2 (23.8x) & 1178.6 (2.8x) & 576.6 (11.3x) & 1827.2 (6.1x) \\
\hline
\end{tabular}
}
\end{table*}

These gains become even more pronounced when evaluated through the hardware-oriented ballpark cost in Table~\ref{tab:ballpark-cost-by-setting}, which indirectly captures the expected workload in hardware fine-tuning.
Using the ballpark cost defined in Section~\ref{subsec:metric}, for SK at $(n = 20, p = 2)$ SAFE with distillation uses only about $5.5\%$, corresponding to a $94.5\%$ reduction in ballpark cost.
This aggregate reduction comes from both a smaller active parameter count, from $268.4$ to $96.9$ on average, and a reduced number of first-hit steps using the exact energy objective, from $59.2$ to $7.6$ on average.
Each setting shows the same pattern: SAFE with distillation lowers the estimated cost relative to exact-only in every SK, 2D grid, and Max-Cut setting.

These convergence and ballpark cost results, together with the final approximation ratio results in Fig.~\ref{fig:results-overview}, show that distillation provides a practical cost reduction within the SAFE workflow.
LWPP pre-training supplies the main advantage by producing useful initial parameters for exact fine-tuning, while distillation often strengthens it by improving the initial exact energy and removing low-magnitude parameters.
SAFE with distillation therefore reduces the burden of exact fine-tuning through two coupled mechanisms: fewer active parameters and a lower first-hit step.

\begin{figure}[!b]
\centering
\includegraphics[width=\columnwidth]{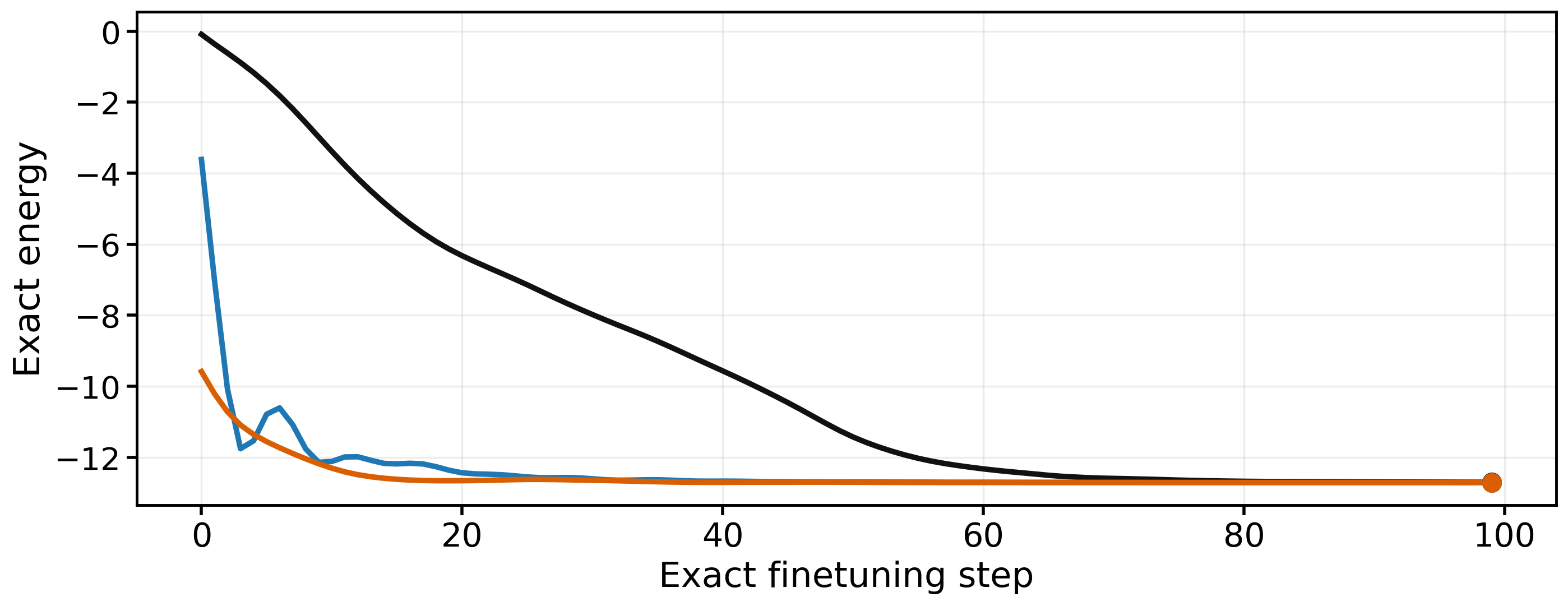}
\caption{Representative exact fine-tuning trajectories for an SK instance ($n=20$, $p=4$). The black curve shows exact-only from the original initialization, while the blue and orange curves show exact fine-tuning initialized with parameters obtained from LWPP pre-training for selected representative runs. Although these runs start from different exact energies, they reach the same low-energy plateau considerably earlier than exact-only, showing that the fast convergence is not explained solely by a lower initial exact energy.}
\label{fig:param-aligned-finetuning}
\end{figure}

\begin{figure*}[!t]
\centering
\includegraphics[width=0.98\textwidth]{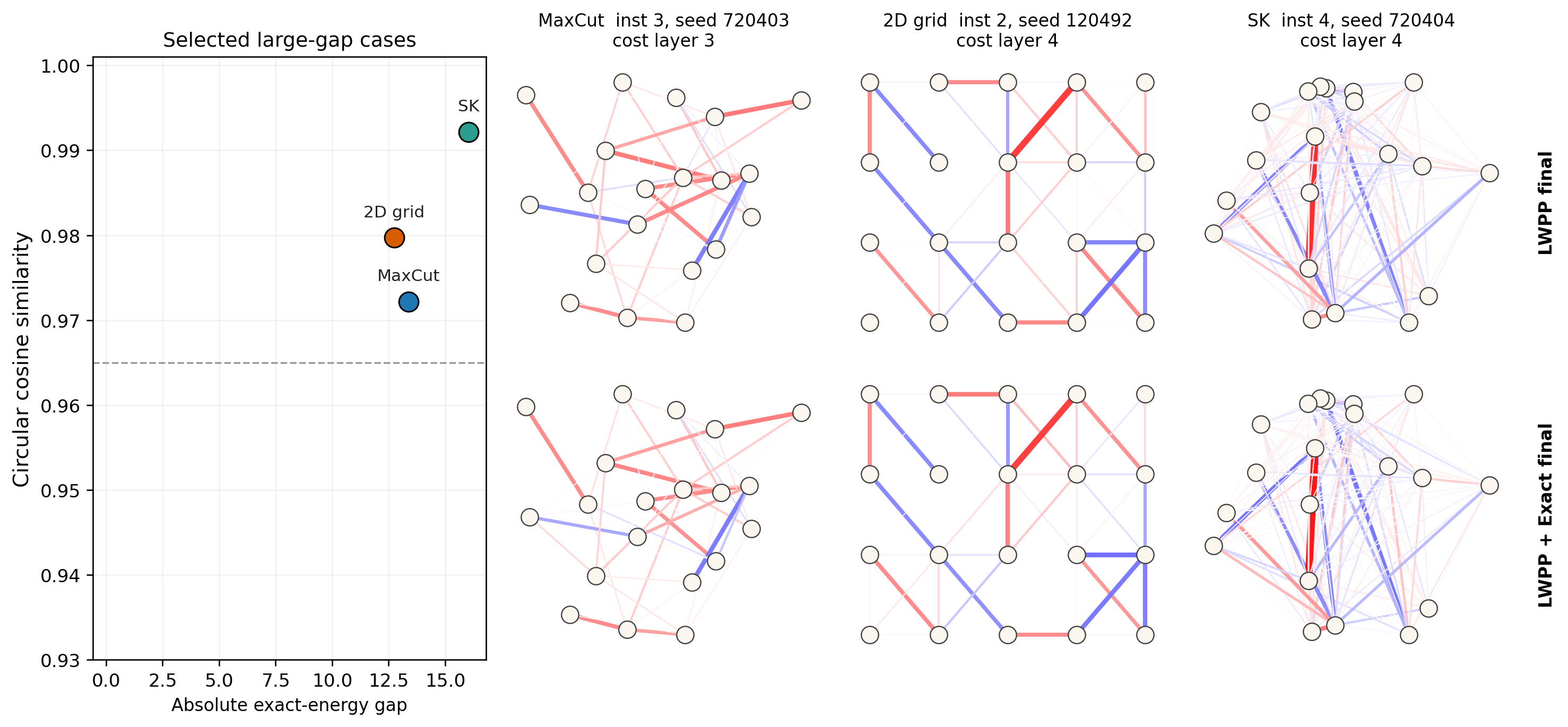}
\caption{Cost-angle structure in selected SAFE cases with large exact-energy changes after fine-tuning. The exact-energy gap is the absolute difference between the exact energy at the LWPP-final parameters and the exact energy after LWPP+exact fine-tuning. The left panel compares this gap with the circular cosine similarity between the corresponding vectors of ma-\qaoa\ cost angles, where each vector collects the trainable angles assigned to the cost-Hamiltonian terms. The graph panels visualize the selected cost layer for each family before and after exact fine-tuning; edge color and width encode the corresponding cost angles, and node layouts are fixed within each column to make structural changes visible.}
\label{fig:maxcut-param-similarity}
\end{figure*}

\subsection{Accelerated Convergence and Preserved Angle Structure}
\label{subsec:param-similarity}

To explicitly demonstrate how the SAFE workflow accelerates exact fine-tuning, Fig.~\ref{fig:param-aligned-finetuning} compares the optimization trajectories of a representative SK instance ($n=20$, $p=4$). The exact-only curve gradually descends and converges to the final energy after many optimizer updates using the exact energy objective. By contrast, the curves initialized with parameters obtained from LWPP pre-training begin exact fine-tuning from different exact energies but approach the same energy considerably earlier. This remains notable even after accounting for the better initial exact energies: the early exact-energy descent remains visibly steeper. Consequently, the surrogate-guided initialization not only provides lower-energy starting points but also places the variables on a steeper descent trajectory toward the low-energy plateau compared to exact-only.

To test whether this steep exact fine-tuning descent reflects preservation of useful parameter structure, we focus on cases where exact fine-tuning changes the exact energy substantially. Here, the exact-energy gap is the absolute difference between the exact energy at the LWPP-final parameters and the exact energy after the subsequent LWPP+exact fine-tuning stage. For selected large-gap cases from Max-Cut, 2D grid, and SK, Fig.~\ref{fig:maxcut-param-similarity} compares this gap with the circular cosine similarity between the corresponding vectors of trainable cost angles. Here, the circular cosine similarity between two angle vectors $\boldsymbol{\phi}$ and $\boldsymbol{\psi}$ of dimension $d$ is defined as $\frac{1}{d}\sum_{k=1}^{d}\cos(\phi_k - \psi_k)$, which equals $1$ when all angles agree modulo $2\pi$ and $0$ for uncorrelated configurations. The right panels visualize the edge-indexed cost angles for one selected layer in each family, with the top row showing the LWPP-final parameters and the bottom row showing the parameters after exact fine-tuning. These visualizations reveal that exact fine-tuning largely preserves the distinct patterns of positive and negative edge angles initially established by the surrogate model.

The selected cases remain strongly aligned in these cost-parameter vectors even when exact fine-tuning changes the exact energy markedly. All three points lie well above the dashed similarity guide in the left panel, which represents the minimum cosine similarity among all cases. The graph visualizations show the same qualitative pattern: the prominent positive and negative edge-angle structures remain recognizable after exact fine-tuning, although individual magnitudes are adjusted. Therefore, for these representative large-gap cases, the exact fine-tuning process operates by refining the meaningful cost-parameter structure obtained from the LWPP surrogate rather than moving to an entirely unrelated parameter set.

This parameter-level analysis helps explain both the rapid convergence and the usefulness of distillation. The trajectory comparison in Fig.~\ref{fig:param-aligned-finetuning} shows that exact fine-tuning initialized with parameters obtained from LWPP pre-training can reach the low-energy plateau considerably earlier than exact-only. The similarity analysis in Fig.~\ref{fig:maxcut-param-similarity} further shows that this fast descent can occur while preserving the cost-parameter structure established during LWPP pre-training. This combined evidence supports the interpretation that the SAFE framework accelerates training by using LWPP pre-training to produce a meaningful, structurally aligned starting parameterization for exact fine-tuning.

\section{Discussion}
\label{sec:discussion}

\begin{figure}[t]
\centering
\includegraphics[width=\columnwidth]{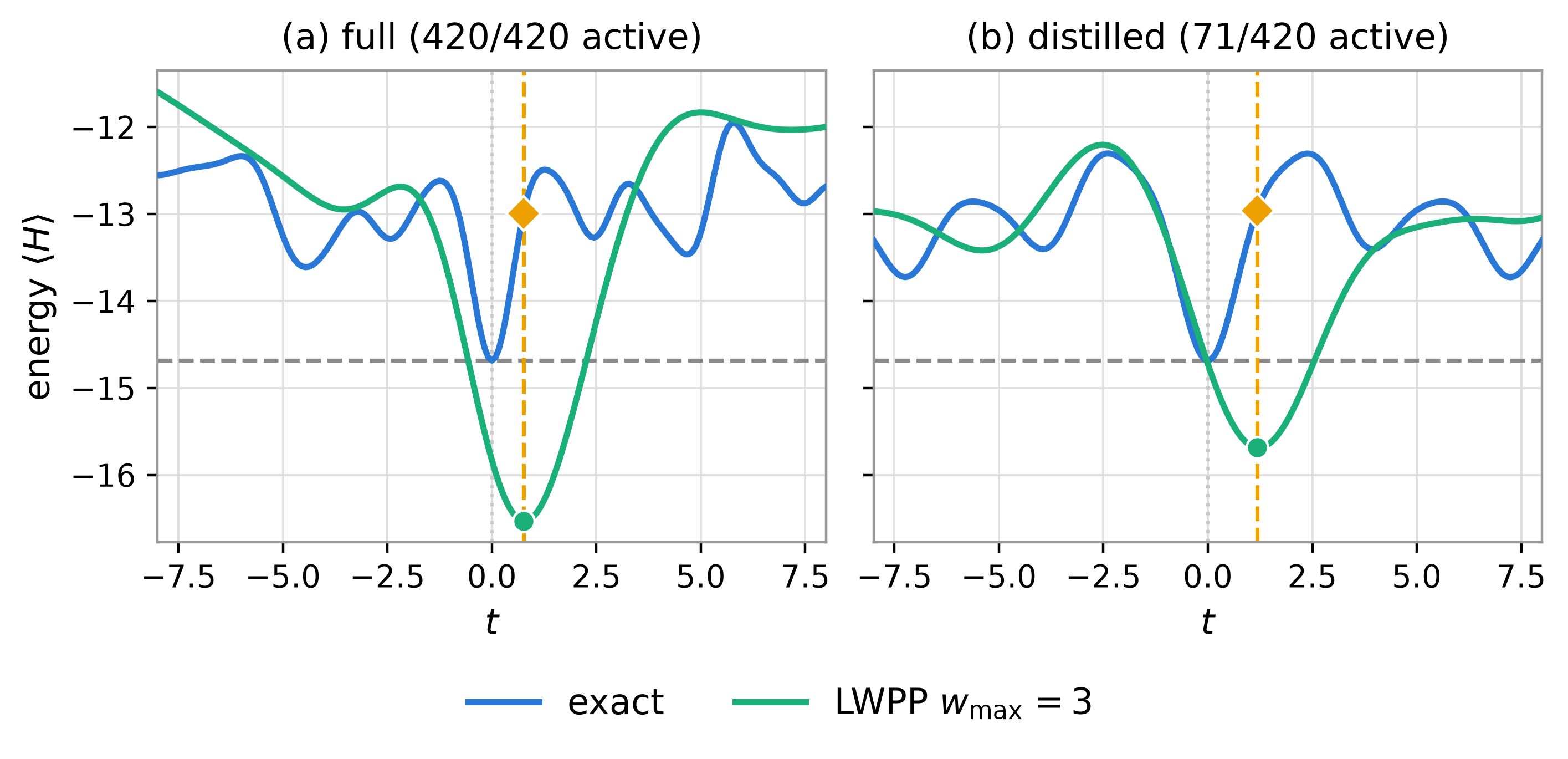}
\caption{Representative one-dimensional exact and LWPP energy profiles for an SK instance ($n=20$, $p=2$, $w_{\max}=3$). Starting from $\boldsymbol{\theta}_{\mathrm{final}}$, which attains the global minimum of the exact objective, all active parameters are displaced jointly along the normalized LWPP-gradient direction by $t$. (a) Full setting with 420 parameters. (b) Distilled setting with 71 active parameters using a cutoff of $0.3$. The vertical and horizontal gray lines mark $t=0$ and the ground-state energy, respectively. The green dot and orange vertical line mark the lowest plotted LWPP value, and the orange diamond gives the exact energy at the same $t$. In both settings, the smoother LWPP profile has a low-energy region near $t=0$.}
\label{fig:fourier-smoothing}
\end{figure}

The observed convergence behavior motivates a closer examination of the local relationship between the LWPP and exact objectives. In particular, we ask whether LWPP retains locally useful structure even when its energy values differ from the exact objective. To examine this question, we evaluate the exact and LWPP energies for the representative instance in Fig.~\ref{fig:fourier-smoothing} along $\boldsymbol{\theta}(t)=\boldsymbol{\theta}_{\mathrm{final}}-t\,\boldsymbol{g}_{\mathrm{LWPP}}/\Vert\boldsymbol{g}_{\mathrm{LWPP}}\Vert$, where $\boldsymbol{g}_{\mathrm{LWPP}}=\nabla E_{\mathrm{LWPP}}(\boldsymbol{\theta}_{\mathrm{final}})$. This is a local directional comparison around the optimum in the energy landscape.

Along both sweeps, the exact landscape fluctuate on a fine scale, whereas the LWPP landscape is smoother. This behavior is consistent with the intuition that weight truncation may suppress higher-frequency variations and retain lower-frequency components. Bbecause LWPP truncates propagated operators, its energy values need not be the exact ground-state energy. The LWPP low-energy region occurs near $t=0$, which corresponds to the global minimum of the exact objective.

For our illustrative instance, the point corresponding to the lowest LWPP value lies on a steep slope of the exact landscape that descends toward the global minimum at $\boldsymbol{\theta}_{\mathrm{final}}$, suggesting that LWPP favors a region with a strong exact-energy descent gradient despite the mismatch between the two objectives. 
The distilled landscape exhibits the same qualitative relationship after parameter distillation: the LWPP landscape remains smooth, its low-energy region remains near $t=0$, and the corresponding exact landscape retains a descent gradient toward $\boldsymbol{\theta}_{\mathrm{final}}$. 

Because the comparison is based on one instance and a single one-dimensional sweep for each setting, this viewpoint should be seen as a specific local observation rather than a general mechanism. 
It also should not be conflated with barren-plateau mitigation. Our experiments show that LWPP pre-training provides useful initial parameters and reduces the number of subsequent exact optimization steps, but we do not analyze the scaling of gradient variance with system size. 
Therefore, the present results establish neither rigorous performance guarantees of LWPP nor the presence of  barren plateaus. Rather, we claim SAFE reduces exposure to potentially difficult exact optimization landscapes by performing the initial search with a classically tractable surrogate and reserving exact optimization for a shorter fine-tuning stage.

We point to three directions for future work. First, all results in this work are obtained from noiseless classical emulators. 
It remains to be established whether the observed reductions in active parameters and exact optimization steps persist under finite-shot, noisy QPU executions. 
These experiments should also include larger instances beyond 30 qubits, where exact emulation is no longer feasible, and broader problem families to assess scalability and practical utility. 
Second, resource savings must be assessed jointly with attained solution quality because even small approximation-ratio differences can incur practical costs. It is possible that further gains at already high ratios require considerably greater resources. 
Future work should therefore compare the resources required to reach fixed approximation-ratio targets across problem classes and instances. 
Third, the interpretation of LWPP as a reliable guide for the exact objective remains empirical. Theoretical analysis should determine when weight truncation preserves the optimization landscape, while formal spectral analysis should test whether it preferentially suppresses higher-frequency components.

\section{Conclusion}
\label{sec:conclusion}
In this work, we present SAFE ma-\qaoa, a surrogate-assisted workflow that combines \lwpp\ pre-training, optional parameter distillation, and exact fine-tuning for ma-\qaoa\ circuits. The main practical benefit of this workflow is that it reduces the quantum resources needed for the exact optimization stage. In particular, SAFE with distillation preserves a strong final approximation ratio while removing a large fraction of trainable angles and reducing the fine-tuning cost relative to exact-only.

We also experimentally demonstrate this benefit. In many cases, the energy obtained after \lwpp\ pre-training remains far from the energy reached after exact fine-tuning. Nevertheless, the corresponding parameters are similar to their exact fine-tuning counterparts. This indicates that \lwpp\ does not need to solve the exact optimization problem directly; it can prepare a parameter already close to the optimum. 
Distillation strengthens this effect by removing angles that remain close to zero, thereby reducing the dimension of the exact optimization without discarding the dominant parameter structure learned during surrogate pre-training.

\section*{Acknowledgment}

The authors thank the Quantum AI Team at NORMA for valuable discussions and support. The authors also thank Hyun-Soo Kim and Junseo Lee for helpful discussions and feedback.

\clearpage
\IEEEtriggeratref{49}
\bibliographystyle{IEEEtran}
\bibliography{reference}

\end{document}